# Secure Routing Protocol to Mitigate Attacks by using Blockchain Technology in Manet

Nitesh Ghodichor[1], Raj Thaneeghavl. V[1], Dinesh Sahu[1], Gautam Borkar[2], Ankush Sawarkar[3]

[1]Department of Computer Science and Engineering, SRK University, Hoshangabad Road, Misrod, Bhopal-462026 (India),
[2]Department of Information Technology, RAIT, DY Patil University Sector 7. Nerul, Navi Mumbai 400706 (India),
[3]Department of Computer Science and Engineering, VNIT, South Ambazari Road, Nagpur, Maharashtra.  440010 (India),

## Abstract

*MANET is a collection of mobile nodes that communicate through wireless networks as they move from one point to another. MANET is an infrastructure-less network with a changeable topology; as a result, it is very susceptible to attacks. MANET attack prevention represents a serious difficulty. Malicious network nodes are the source of network-based attacks. In a MANET, attacks can take various forms, and each one alters the network's operation in its unique way. In general, attacks can be separated into two categories: those that target the data traffic on a network and those that target the control traffic. This article explains the many sorts of assaults, their impact on MANET, and the MANET-based defence measures that are currently in place. The suggested SRA that employs blockchain technology (SRABC) protects MANET from attacks and authenticates nodes. The secure routing algorithm (SRA) proposed by blockchain technology safeguards control and data flow against threats. This is achieved by generating a Hash Function for every transaction. We will begin by discussing the security of the MANET. This article's second section explores the role of blockchain in MANET security. In the third section, the SRA is described in connection with blockchain. In the fourth phase, PDR and Throughput are utilised to conduct an SRA review using Blockchain employing PDR and Throughput. The results suggest that the proposed technique enhances MANET security while concurrently decreasing delay. The performance of the proposed technique is analysed and compared to the routing protocols Q-AODV and DSR.*

## 1. Introduction

MANET is a wireless network [1] without a central administration. MANET's simplicity makes it popular. In a MANET, nodes communicate using radio frequency. MANET's open design is vulnerable [1]. The attacker may target the network while data or packets are transferred or by monitoring network activity. Attacks might target data or control. After network control information is delivered, MANET sends data in packets. Attacks are any activity or threat that compromises a network's security. Critical MANET security. Open media prohibits end-to-end linkages. Security is maintained via many techniques. Key-based protocols and IDS authenticate Plan. SRA and Blockchain are protected by Secure Routing Protocol. Blockchain ensures data record security and validity without a third party. Blocks store data in a blockchain. When full, blocks are sealed and added to the chain.

Blockchain allows decentralised trust and security. New blocks are added sequentially. They're added to blockchain's "end." Once a block is added to the blockchain, changing its contents requires a majority vote. Each block has a hash, the previous block's hash, and the date. Hash codes convert digital data to numbers and letters. Data changes the hash code.

A cryptographic hash function is a mathematical operation [3]. Standard hash functions accept variable-length inputs and produce fixed-length [2] outputs (hashes). Cryptographic hash functions combine message-passing and security [3]. Traditional hash functions are secured by cryptographic hash functions [3], which makes it more difficult to decipher message content or recipient information. Given the network's vulnerability, assaults are classed as follows. SRA with blockchain enables network security by producing hash codes for each transaction and authenticating each node for data and control packets.

Early research focused only on secure data transmission routing mechanisms, even though attacks might occur in any MANET security layer. Most MANET security approaches require network nodes to perform a disproportionate amount of work based on constantly updated topology control. This wastes resources and puts users at risk of hacker assaults. Without verification, nodes distrust each other. Attacks that swiftly enter and change residential nodes can harm network performance. This paper presents a blockchain-based, management-secure, QoS-enhanced MANET.

By combining a hash function and blockchain technology to authenticate nodes that join the network, SRA with blockchain maintains the network's security. Smart network nodes keep track of the surrounding node's characteristics and if it makes a substantial contribution. The decision is made using the delay criterion. The suggested algorithm offers precise data for decision-making and maintain network security, the node is removed from the network based on its delay factor.

## 2. RELATED WORK

U. Srilakshmi et al in January 19, 2022, The "A Secure Optimization Routing Algorithm for Mobile Ad Hoc Networks" proposed algorithm [4] uses optimization algorithms, which find new ways to progress routing, to provide trust-based safe and energy-efficient navigation in MANETs. Once the fuzzy clustering approach has been activated, the Cluster Heads (CHs) are selected based on the indirect, direct, and recent trust values that each CH carries. Based on trust ratings, value nodes were also found. The CHs engage in multi hop routing using the projected protocol, which selects the best routes depending on latency, performance, and connection within the boundaries of the course. [4]

R. Thiagarajan and R Ganesan, a system for detecting malicious nodes at each destination was proposed in Optimised with Secure Approach in Detecting and Isolation of Malicious Nodes in MANET in 2021. Following the detection procedure, it is isolated and discarded while the path is established using other techniques. Paths for a group of disparate nodes can be found with the aid of an algorithm that enables multipath reliable routing. The pathways are reordered based on the reliability index. [5]

In their article published in March 2020, Neenavath Veeraiah and B. T. Krishna offer a method for safe multipath routing and intrusion detection in MANET [22]. The approach uses an optimization algorithm as its foundation offer MANET with effective multipath routing. In order to effectively address the energy and security problems in the MANET, the cluster head (CH) selection and intrusion detection algorithms, also known as fuzzy clustering and fuzzy Naive Bayes, are utilise (fuzzy NB). The bird swarm-whale optimization algorithm, also known as BSWOA, it is then utilised in order to further multipath routing by utilising secure nodes. BSWOA is a combination of the bird swarm optimization algorithm (BSA) and the whale optimization algorithm (WOA). [6,7,21]

Deepak Sinwar, Nisha Sharma, DSDV, AODV, and AOMDV discussed an ant colony optimization algorithm based on shortest path in MANET in 2020. Ant-Colony Optimization (ACO) and Particle Swarm Optimization are two Swarm Intelligence (SI)-based methodologies that have been used to address the demand for an optimum path for communication among nodes (PSO). The protocol must be optimised to choose the best way in order to extend the length of communication availability. These many routing issues can be resolved using SI-based techniques. ACO offers the best packet delivery speed, throughput, power efficiency, and packet delay. [8]

In 2018, G.M Borkar and A.R Mahajan introduce "The Security Aware Dual Authentication Based Routing (SDAR) and Highly Efficient Dual Authenticated Routing (HEDAR). It uses Fuzzy logic control (FLC) based prediction algorithm and game theory. Cipher text policy attribute-based encryption (ABE) is used for secure transmission. Evaluate several assessment measures of node density levels. [9,19,27]

Nada Mouchfiq, in 2019 explains network security using the Blockchain to enable the adoption of new processes, enable the exchange of messages and information, and enable autonomous device coordination. they put up a security approach called "MPR Blockchain" it is based on the Blockchain and is more suited to our needs as a group working in ad hoc networks. [10]

David Cordova, Alexandre Laube explain in 2020, blockgraph: A blockchain for mobile ad hoc Networks, which we refer to as "blockgraph," discusses the difficulties in adopting a DAG-based blockchain for MANETs. We describe blockgraph features. This contains a consensus mechanism that is resilient to network partitions, blockgraph protocol specifications that maintain the blockgraph data structure, and a group management system that responds to network topology changes and provides the block graph framework with topological information. [11]

Masood uses blockchain to illustrate reputation-based routing in MANET. The difficulty level and score are then combined to determine a node's reputation. The quickest, most reputable path between a source and destination node is determined using this reputation as part of a novel routing metric. By barring malicious nodes from participating in packet routing, the intention is to deter them from being malicious. In the context of routing threats, a combined simulation of the Blockchain and routing algorithm demonstrates an improvement in overall packet delivery [12].

Bhagyalakshmi proposed Q-AODV [13]: A Flood Control Ad-Hoc on Demand Distance Vector Routing Protocol known as Q-AODV seeks to reduce the number of intermediary nodes involved in the route discovery process in order to minimize the overall volume of control packets forwarded by network nodes. This is accomplished by using the queue length of the node to control the route request (RREQ) broadcast storm. At each intermediary node, the queue vacancy proportion is compared with the random number that the source appends with RREQ. The intermediate node broadcasts the RREQ packet if the randomly generated number is less than the proportion of queue vacancies. Decreasing the number of crowded nodes forwarding the RREQ packets, lengthens the network's total lifespan and enhances QoS metrics. The proposed algorithm Q-AODV is an enhancement to AODV that searches for a less congested route depending on queue vacancies. The proposed technique QAODV somewhat lowers jitter, average end-to-end delay, and throughput compared to AODV [13,17,18].

N. Prasath and J. Sreemathy. The performance of the optimized dynamic source routing protocol (DSR) for MANETs is examined in the 2018 article, Describe Optimized dynamic source routing protocol for MANETs. The Firefly algorithm is used to modify the conventional DSR algorithm in order to determine the best routes between the communication nodes. In order to increase DSR routing performance with well-organized packet transfers from the source to the destination node, the suggested technique on MANET uses the Firefly algorithm. Based on link quality, node mobility, and end-to-end delay, the best path is discovered. [14]

Ravilla Dilli, 2016, discusses Security features in MANETs using the Secured Hash Algorithm (SHA3-256) for secured routing in Mobile Ad hoc Networks' hybrid routing technique (MANETs). To ensure data integrity and authenticity, he employed the Hashed Message Authentication Code (HMAC). Zone Routing Protocol was the hybrid routing method employed (ZRP). [15]

In January 2015, B. Madhusudhanan et al. created a mobility-based key management technique for mobile ad hoc networks' multicast security. A node's anticipated stability index, which is based on link availability and mobility, is used to first group them. Every weak node in a multicast tree has a strong parent node and vice versa. To send multicast data, a session key-based encryption mechanism is used. Regularly, the initiator node conducts the rekeying process. Because the rekeying interval is a constant dependent on the node type, this technique drastically minimises the rekeying overhead. [16]

## 3. PROPOSED SCHEME

As there will be many nodes connected to the network, QoS will affect the delay factor. When addressing the problem of the network attack, this study cannot ignore the factor of delay. The most important factor in determining the effectiveness of an ad hoc network is delay. The data takes longer to reach the goal node than expected. Multiple linked nodes and network mobility are the main causes of delay in the intelligent world. The suggested plan guarantees network security while lowering the latency factor.

A huge number of nodes are used to calculate the delay. The number of connections and the length of pauses between the nodes (devices) will reduce latency [1]. Smart network nodes store information about nearby nodes and base their decisions on the outcomes. Nodes keep an eye on whether the nearby device nodes provide a significant contribution. The criteria of delay are used to make the choice. The actual node is removed from the network following the estimated time after reviewing the surrounding node's response time. This also employs an algorithm to provide accurate information for making decisions. Before the target node is reached, the entire process will be looped.

Algorithm: Reducing the mobile ad hoc network delay value to a minimum using the SRABC scheme

Input: Node parameter
Output: Reduce the node parameter function(delay)
Step I: A network with separate mobile nodes and parameters is specified.
Step II: Source and destination node configuration as Sr and Dt.
Step III: Using the AODV protocol, build routes between source and destination. The route is described as Sr, R1, R2, R3... Rn, Dt.
Step IV: For a=1 to n (repeat steps IV to VIII)
Step V: RList find adjacent (a)
For b-1 to length (RList)

{
Step VI: Parameter = Packet
        Delay (b)
Step VII: Using Fuzzy Logic for parameter fuzzification
Step VIII: If the parameter Delay =low
{ Set Rlist(a) as the next communicating node }
}
Step IX: Go to the following next node.
Step X: Finish

Using the cutting-edge algorithm that successfully reduces the delay factor for the significant number of blockchain technology nodes that are connected. The effectively avoid attacks, blockchain technology has suggested the development of a revolutionary Relinquished Blockchain Based Integrity System then to determine the network system's dependability.

## 4. PROPOSED SCHEME

This section offers a thorough explanation of the execution of this suggested system's findings, performance, and comparison methodologies.

### 4.1. A Test Based Setup

This study was implemented in Python using the following system specifications and simulation results. Windows Python 3.7 4GB RAM, Intel Core.

### 4.2. Performance Metrics

BCR's simulation is evaluated using metrics such as the Packet delivery ratio, average throughput, end-to-end delay (number of dropped packets), routing overhead (latency in acquiring routes), block height (deposit address), minimum difficulty (number of nodes), movement speed (transmission range), simulation time (latency in generating blocks), and so on When a route request has expired, Blacklisted addresses, a route selected, a timer for the blacklist, and the validity of the route offer are all included in this section.

#### 4.2.1. Packet Delivery Ratio

The ratio of the total number of packets [20] transmitted to the total number of packets delivered from the source node to the network destination node is known as the packet delivery ratio (PDR). It is advantageous to send every data packet to its intended location. The PDR value also improves the effectiveness of the network.

$$PDR = \left( \frac{\sum Received\ Packets}{\sum Sent\ Packets} \right) \times 100$$

#### 4.2.2. Throughput

The receiver receives a specific number of data packets in a given amount of time. Typically, it is calculated in bits per second.

$$Throughput = \frac{\sum Number\ of\ Packets_{received}}{\sum Total\ Time}$$

**4.2.3. End-to-End Delay**

End-to-end delay sometimes referred to as one-way delay, is the length of time it takes a packet to move from source to destination over a network. Widespread in use, in-network surveillance differs from round-trip time RTT in that just one path is measured from source to destination.

$$d_{end\,to\,end} = N\,\frac{L}{R}$$

**4.2.4. Routing Overhead**

The number of packets sent for maintenance and path discovery can be considered.

$$Routing\,Overhead = \frac{Sum\,of\,routing\,packets}{Sum\,of\,data\,packets}$$

**4.2.5. Route Acquisition Latency**

This is the amount of time that passes between a source sending a request/discovery packet to a destination to determine the path and receiving the ensuing first response. The first transmission sending period has been used for the latency measurement if a path demand needs to be retransmitted because it was timed.

**4.2.6. Block Height**

The block height of a particular block determines how many blocks must come before the blockchain. The genesis block was referred to as a blockchain's very first block. Since there are no preceding blocks in the blockchain, there is a zero block height. The length of the blockchain is less than one is frequently used to quantify the cumulative block-chain height of the most recent block. The biggest block is on the blockchain.

$$Block\,height = block\,chain - 1$$

**4.2.7. Deposit Address**

A deposit address is a transitional address on a native network of cryptocurrency used to connect deposits on a Gate Hub platform with individual user wallets. The user is not included in the registry addresses created by Gate Hub (for cryptocurrencies other than XRP).

**4.2.8. Maximum Difficulty**

The length of time it takes miners to add fresh blocks of transactions to the blockchain is controlled by a quantity called the difficulty.

$$Maximum\,difficulty = \frac{\max\,imum\_t\arg et}{1} = \infty$$

**4.2.9. Number of Nodes**

A blockchain is created from data blocks. A system that has a complete copy of the blockchain transaction history is referred to as a complete node.

**4.2.10. Simulation Time**

The amount of time it takes for a system or process to complete its whole execution, including all of the internal sub-processes. The measurement is in seconds.

**4.2.11. Block Generation Latency**

The time it takes to construct the following block of transactions in the chain is the network latency, commonly referred to as "block time." In other words, it refers to the amount of time a user must wait after clicking the transmit button before seeing their transaction appear on the blockchain.

**4.2.12. Simulation Outputs**

The following figures are an analysis of the simulation output. Create the blockchain network first at random. After then, the blockchain is built in the MANET, the node is identified by its pertinent weight, and communication takes place within the MANET. Additionally, the transaction has been completed in MANET by determining the shortest path for communication. Finally, a variety of assaults are taken into account in the dataset during training, and a variety of attacks in the dataset have been verified through training and testing.

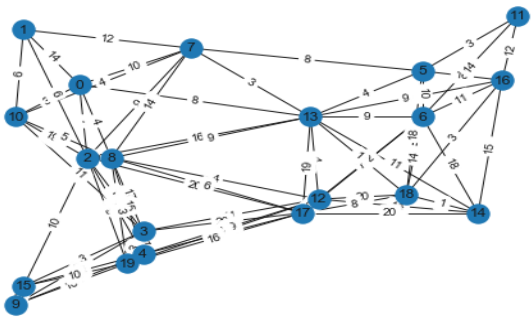
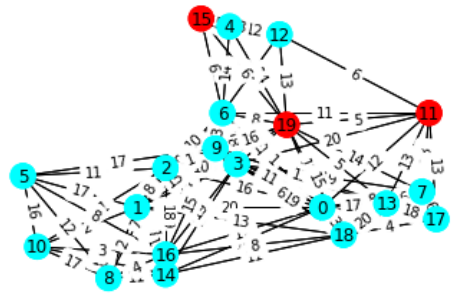

Figure 1. Node creation in block-chain.          Figure 2. Shortest path calculation.

Figure 1, depicts the establishment of a node with the appropriate weight in the MANET block-chain. Following the creation of the block and hashing of the data, a node is constructed using a blockchain, with each node having the proper weight. The determination of the shortest way to deliver the data from one node to the following node is shown in Figure 2. The shortest path will be from node 11 to node 19 and from node 19 to node 15, where node 11 is the source node and node 15 is the destination node. If node 12 is used, it will take two nodes to get there.

## 4.3. Simulation Output without Blockchain in MANET

The following figures show the simulation outcome without a blockchain. Without taking into account blockchain metrics like throughput, average end delay, packet delivery ratio, and routing overhead, the following result was calculated for 5 nodes and 100s of time.

The throughput against time without blockchain security is shown in Figure 3. According to this finding, throughput in the 20s is 400 packets, it progressively climbs to 700 packets in the 40s, reaches 950 packets in the 60s, 1590 packets in the 80s, and 1780 packets in the 100s.

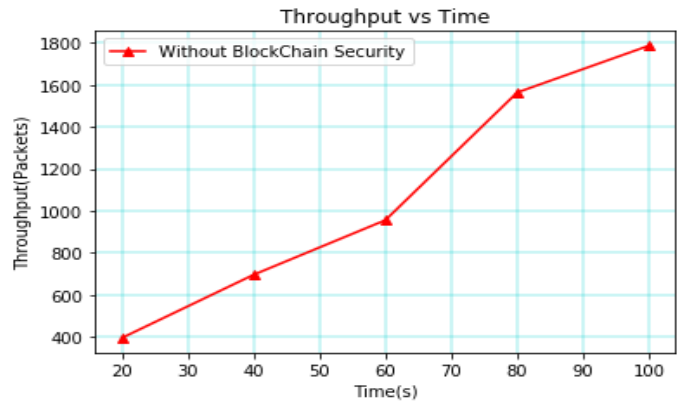

Figure 3. Throughput vs time for without blockchain security.

The average end latency over time without blockchain is shown in Figure 4. This graph shows that when the pause duration increases over time, the average end delay does as well. When the pause time reaches 100, the average end delay is 22.0 seconds. Also indicates that as the nodes expand to five, the end-to-end delay also increases and reaches one millisecond (ms) without blockchain.

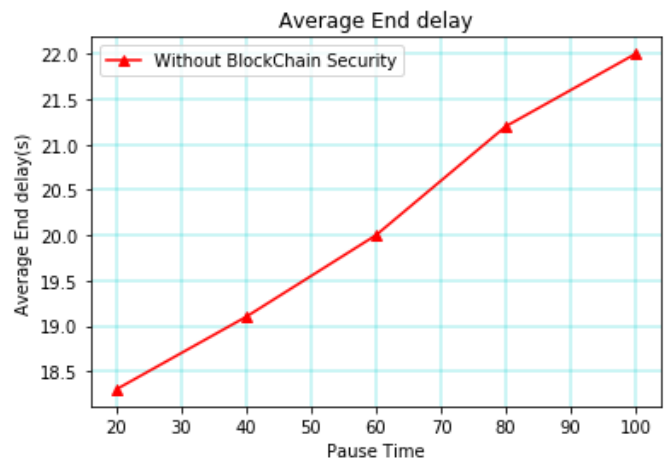

Figure 4. Average end delay vs time for without Blockchain.

The packet delivery ratio over time for no blockchain is shown in Figure 5. This finding indicates that as time increases effectively, the packet delivery ratio also rises and reaches 89.5 percent for 100s.

The routing time versus overhead for no blockchain is shown in Figure 6. Without blockchain, this graph shows that the latency grows to 100s and the routing overhead reaches 69 percent.

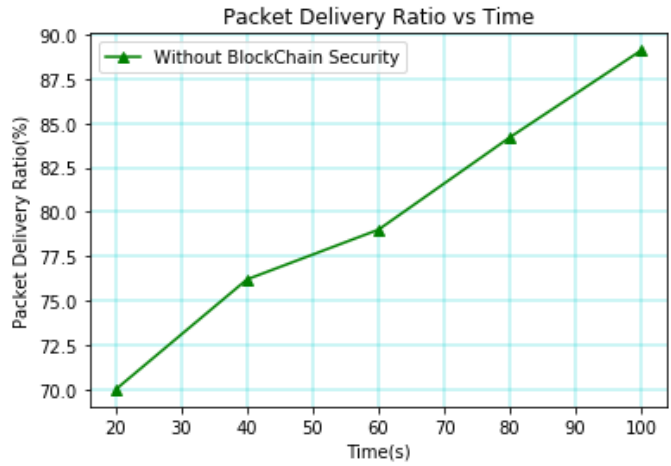

Figure 5. Packet delivery ratio vs time for without Blockchain.

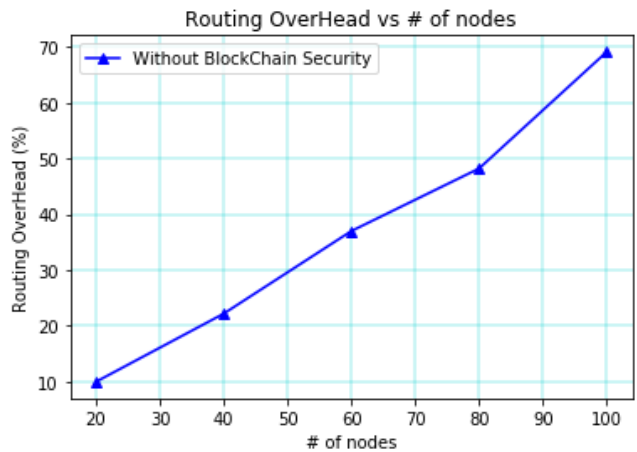

Figure 6. Routing overhead ratio vs time for without Blockchain.

Table 1, depicts the parameters vs time without blockchain in MANET. Throughput, average end delay, packet delivery ratio, and routing overhead [25] were taken into consideration for the parameters in that table, which took into account the time between 20s and 100s. When the time is 20 seconds, the throughput reaches 400 packets, the average delay reaches 18.3 seconds, the packet delivery ratio reaches 70%, and the routing overhead reaches 10%. If the time is 40s, then the throughput attains 700 packets, the average delay attains 19.1s, the packet delivery ratio achieves 76 percent, and the routing overhead attains 23 percent.

Table 1. Parameters vs time without blockchain in MANET.

| Time (s) | Throughput (packets) | Average end delay(s) | Packet delivery ratio (%) | Routing overhead (%) |
|---|---|---|---|---|
| 20 | 400 | 18.3 | 70 | 10 |
| 40 | 700 | 19.1 | 76 | 23 |
| 60 | 950 | 20 | 79.5 | 38.4 |
| 80 | 1590 | 21.3 | 84 | 49 |
| 100 | 1780 | 22 | 89.5 | 69.8 |

If the time is 60s, then the throughput attains 950 packets the average delay attains 20s, packet delivery ratio achieves 79.5 percent, and the routing overhead attains 38.4 percent. If the time is 80s, then the throughput attains 1590 packets, the average delay attains 21.3s, the packet delivery ratio achieves 84 percent, and the routing overhead attains 49 percent. If the time is the 100s, then the throughput attains 1780 packets, the average delay attains 22s, the packet delivery ratio achieves 89.5 percent, and the routing overhead attains 69.8 percent. Table 2 shows the parameters vs nodes without blockchain in MANET.

Table 2. Parameters vs node for without blockchain.

| Node | End to delay (ms) | Packet delivery ratio (%) | Routing overhead (%) |
|------|-------------------|---------------------------|----------------------|
| 1 | 0.2 | 20 | 14.6 |
| 2 | 0.38 | 30 | 21 |
| 3 | 0.59 | 40 | 28.8 |
| 4 | 0.8 | 50 | 35 |
| 5 | 0.96 | 60 | 43.6 |

For MANET without blockchain, table 2 above compares the parameters and nodes. Variables including average end delay, packet delivery ratio, and routing overhead were taken into consideration in addition to the 5 nodes. In the case when the node is 1, the end-to-end delay is 0.2 milliseconds, the packet delivery ratio is 20%, and there is a 14.6 percent routing cost. The end-to-end delay, packet delivery rate, and routing cost are all 0.38 milliseconds, 30%, and 21%, respectively, when the number of nodes is increased to 2. The end-to-end delay, packet delivery rate, and routing overhead all decrease to 0.59 milliseconds, 40%, and 28.8% respectively, when the number of nodes is increased to 3. The end-to-end delay increases to 0.8 ms, the packet delivery rate increases to 50%, and the routing overhead reach at 35% when the number of nodes is increased to 4. The end-to-end delay, packet delivery rate, and routing overhead all increase to 0.96 ms, 60%, and 43.6%, respectively, when the number of nodes is increased to 5.

## 4.4. Simulation Output with Blockchain in MANET

In the following graphics, the simulation output using blockchain in MANET is discussed. The results are shown in the figures below for relevance weight, model accuracy, model loss, confusion matrix, and security level.

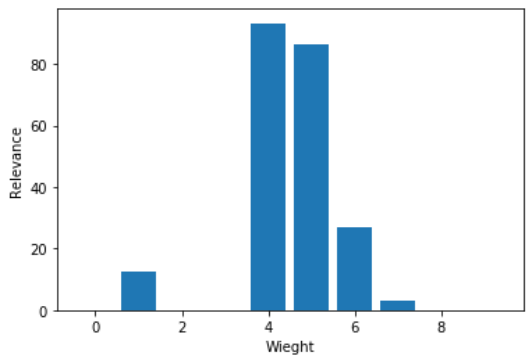
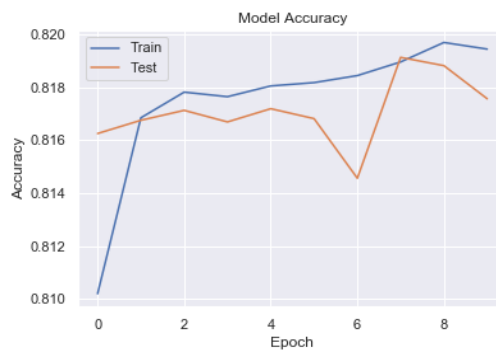

Figure 7. Relevance vs weight.                Figure 8. Model accuracy of the dataset.

The relationship between relevance and weight is depicted in Figure 7. According to relevance, the weight will progressively increase before decreasing. The data set's model accuracy is shown in Figure 8. The accuracy and epoch are plotted using the training dataset and the test dataset. When compared to the test dataset, the training dataset's accuracy will be very high.

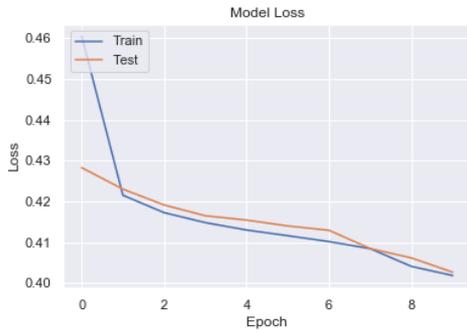

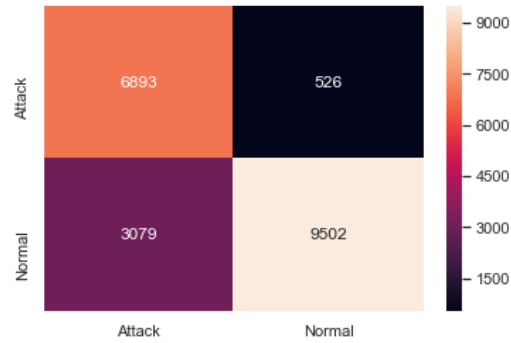

Figure 9. Model loss of dataset.          Figure 10. Confusion matrix for classification.

Figure 9 shows the relationship between the loss and the period. The loss relative to the test dataset will be very little as the training dataset is gradually reduced. And the confusion matrix for classification is displayed in Figure 10. The dataset takes into account a variety of attacks, including Black hole [18], Grey, Warm, Sybil, Altered/Replayed, Spoof Routing, Sinkhole, Hello Flood, and DOS-Apache2 attacks. The total dataset will then be 10,028 in this approach. After 6893 datasets successfully identify the attack, 3079 datasets are incorrectly categorised. Attacks or normal behaviour are unpredictable by 526 datasets. Finally, 9502 datasets successfully categorize attacks that are normal or malicious.

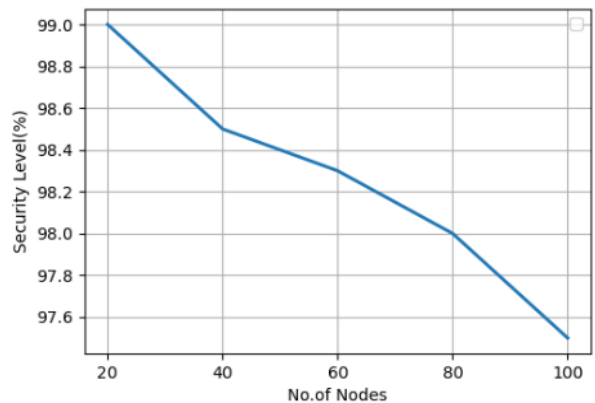

Figure 11. Security Level in Blockchain.

Figure 11 shows the relationship between the blockchain's security level and the number of nodes. The graph shows that the security level will reach a maximum of 99 percent if the number of nodes is modest. The security level slowly declines and reaches 97.5 percent as the number of nodes rises.

## 4.5. Comparison Strategies

Ad hoc On-Demand Distance Vector (AODV) [17,18], Queue Ad hoc On-Demand Distance Vector (QAODV), Dynamic Source Routing (DSR), Secured Encryption Technique with Optimum Route Discovery (SETORD), and Highly Efficient Dual Authenticated Routing (HEDAR) schemes are some of the algorithm schemes with which the performance of the proposed method is compared in this section. The following diagram shows how the comparison process works.

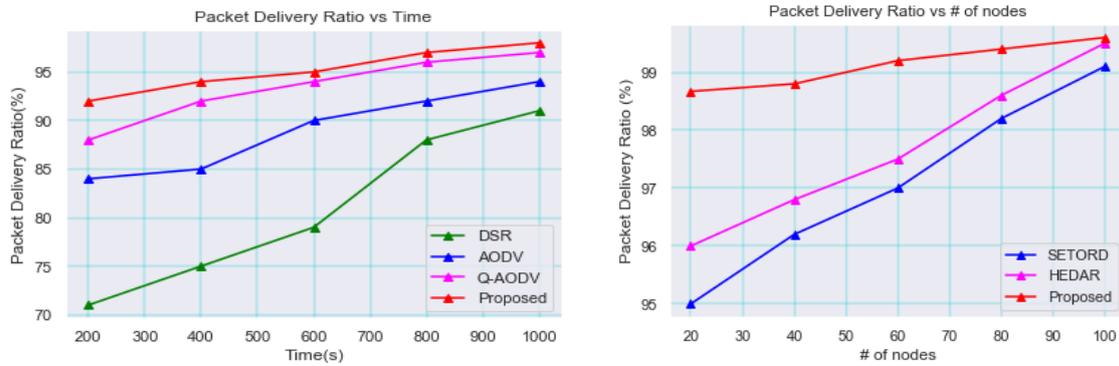

Figure 12. Comparison for Packet Delivery Ratio.

When comparing the packet delivery ratio with earlier approaches like DSR, AODV [17,18], and QAODV, the resulting plot is shown in Figure 12. The graphic makes it obvious that the new strategy has a high packet delivery ratio when compared to the earlier approaches. This successfully implemented strategy lowers packet loss.

When the proposed method was compared to earlier technologies like SETORD and HEDAR, Figure 12 shows the comparison plot between packet delivery ratio and the number of nodes. The chart makes it obvious that, as compared to the earlier method, the Packet delivery ratio reaches its maximum.

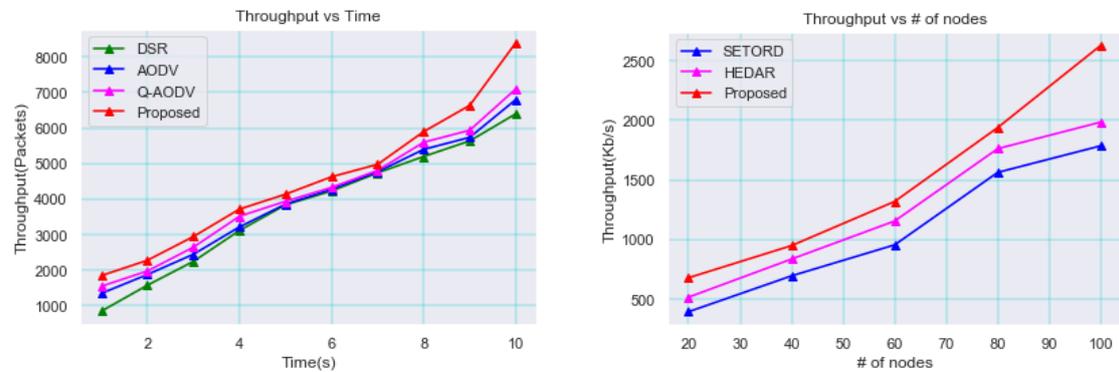

Figure 13. Comparison for Throughput

The results of the comparison between the prior and the proposed method in terms of throughput in packets are shown in Figure 13. The proposed method assures a high delivery rate because it has a high throughput when compared to earlier technology.

While the proposed method was compared to earlier technologies like SETORD and HEDAR, Figure 13 shows the comparison plot between Throughput and the number of nodes. The figure makes it obvious that throughput increases in comparison to the previous technique.

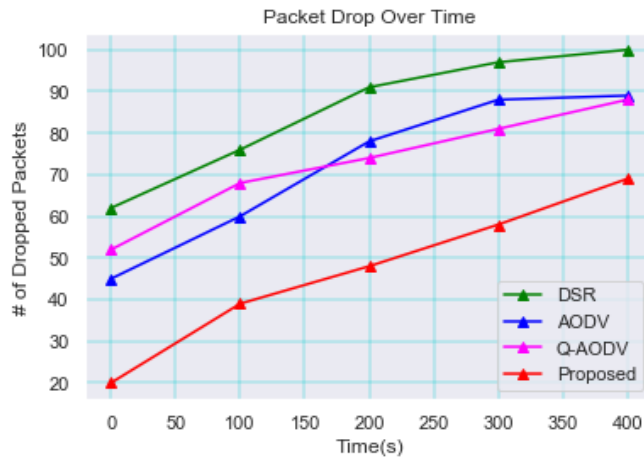

Figure 14. Comparison for Packet Dropped.

Figure 14 shows comparative results between the methods used in the past and those suggested to determine how many packets were dropped in the MANET. The resulting graph demonstrates that the suggested approach does indeed achieve the lowest packet drop rate.

Figure 15 illustrates the end-to-end latency comparison between the proposed network and the prior networks for data packets in the MANET. Thus, the graph shows improved performance for the packets' end-to-end delay in the suggested technique.

Figure 15 illustrates the comparison between the proposed technique and earlier methods like SETORD and HEDAR in terms of the end-to-end delay and number of nodes. The figure makes it evident that the end-to-end delay is ineffective when compared to the earlier technique.

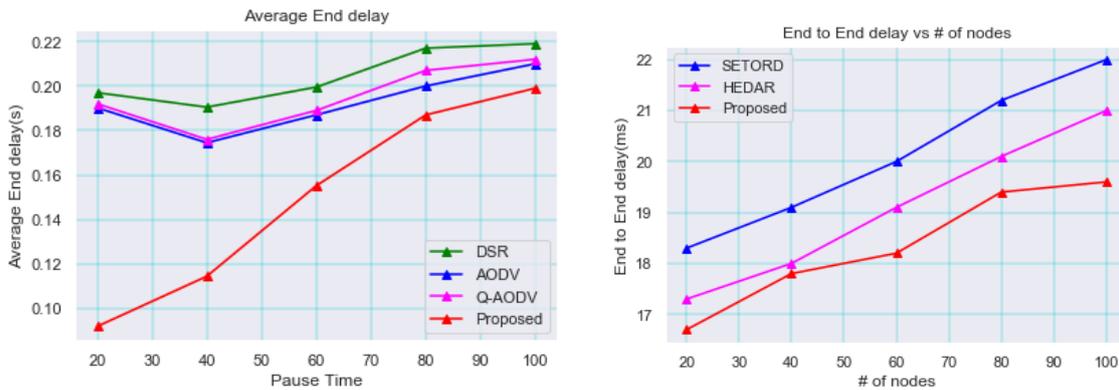

Figure 15. Comparison for End to end delay.

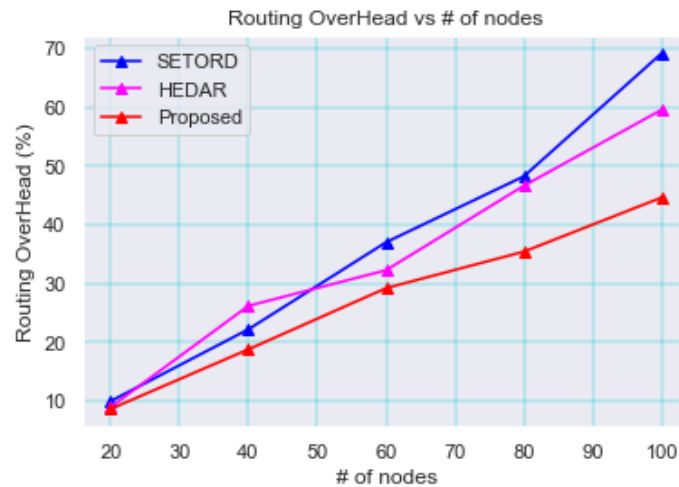

Figure 16. Comparison for Routing overhead.

The resultant plot between the old ways and the new method of routing overhead [26] in the MANET is shown in Figure 16. The graph demonstrates that the suggested method achieves reduced routing overhead in the blockchain MANET as a result.

The parameters for a previous approach and the proposed method are contrasted in Table 3. When compared to older methods like SETORD, Which has a packet delivery ratio of 99.1 percent, and HEDAR, which has a packet delivery ratio of 99.5 percent, the characteristics also vary depending on the number of nodes used. When compared to SETORD's 1787 kb/s and HEDAR's 1986 kp/s, the throughput is 2650 kb/s. End-to-end delays measured using the suggested approach are 19.5 ms, SETORD is 22 ms, and HEDAR is 21 ms. Finally, the routing overhead of SETORD reaches 69.8%, HEDAR reaches 59.9%, but the suggested technique reaches 44.2 percent.

Table 3. Comparison table for Packet delivery ratio, Throughput, end-to-end delay, and Routing Overhead.

| Methodologies | Number of Nodes | Packet Delivery Ratio (%) | Throughput (kbps) | End-to-end Delay (ms) | Routing Overhead (%) |
|---|---|---|---|---|---|
| SETORD | 20 | 95 | 395 | 18.3 | 10 |
| | 40 | 96.2 | 696 | 19.1 | 21.2 |
| | 60 | 97 | 956 | 20 | 38.3 |
| | 80 | 98.2 | 1564 | 21.2 | 47.9 |
| | 100 | 99.1 | 1787 | 22 | 69.8 |
| HEDAR | 20 | 96 | 517 | 17.3 | 9 |
| | 40 | 96.8 | 837 | 18 | 28.4 |
| | 60 | 97.5 | 1156 | 19.1 | 32.5 |
| | 80 | 98.6 | 1763 | 20.1 | 45.2 |
| | 100 | 99.5 | 1986 | 21 | 59.9 |
| Proposed | 20 | 98.6 | 650 | 15 | 8.9 |
| | 40 | 98.8 | 960 | 17.8 | 19.6 |
| | 60 | 99.1 | 1350 | 18.2 | 30.1 |
| | 80 | 99.3 | 1980 | 19.1 | 36.5 |
| | 100 | 99.6 | 2650 | 19.5 | 44.2 |

Thus, when compared to earlier methods, the MANET method for blockchain technology provides more efficiency. Table 4 shows the comparison between the existing method and the proposed method with various parameters calculated.

Table 4. Comparison of an existing method with the proposed method.

| Methodologies | Packet delivery ratio (%) | Throughput (Packets) | Number of packets dropped | Average end to end delay(s) |
|---|---|---|---|---|
| DSR | 91 | 6500 | 100 | 0.219 |
| AODV | 94 | 6800 | 89 | 0.21 |
| QAODV | 96 | 7000 | 87 | 0.208 |
| Proposed | 98 | 8500 | 69 | 0.20 |

According to table 4 above, the suggested technique would achieve a packet delivery rate of 98 percent, a throughput of 8,500 packets, 69 packets that are dropped in the proposed system, and an average end-to-end delay of 0.20 seconds.

Figure 17 shows the results for the QoS settings. The throughput and packet delivery ratio both increased, which led to better efficiency. The smallest end-to-end delay and least amount of dropped packets show improved performance of the suggested strategy. Overall, the suggested method suggested effective QoS Outputs.

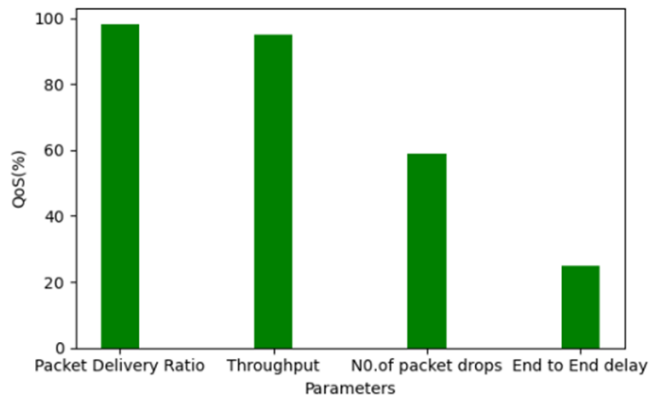

Figure 17. Percentage of QoS for various QoS parameter.

Figure 18 displays attack detection for a variety of assaults, including Blackhole [17], Grey, Warm, Sybil, Altered/Replayed or Spoofed Routing, Sinkhole, Hello Flood, DOS-Apache2 Attack, R2L, and Probe Attacks.

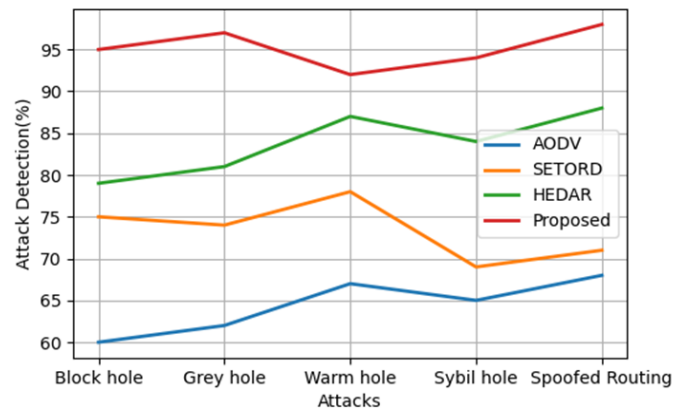

Figure 18(a). Attack detection for various attacks.

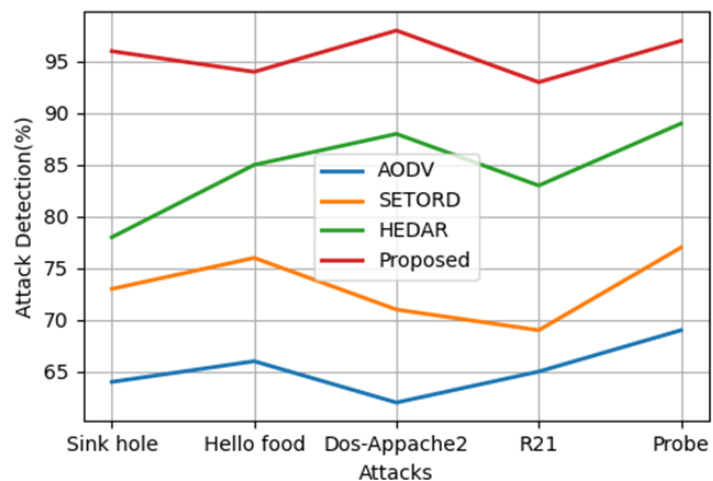

Figure 18(b). Attack detection for various attacks.

Thus, the graph shows that when the proposed method is compared to the prior method, the black hole attack is detected with 95% accuracy, the grey hole attack with 97% accuracy, the warm hole attack with 92% accuracy, the Sybil hole attack with 94% accuracy, the spoofed routing attack with 98% accuracy, the sink hole attack with 96% accuracy, the hello food attack with 94% accuracy, and the DOS attack-Apache2 attack with 98% accuracy. Table 5 shows the comparison between the prior methods and the proposed method for Attack detection (%). Table 5 depicts a comparison between the prior methods and the proposed method for attack detection with percentages.

Thus the graph reveals that the proposed method attains 95% of attack detection when compared with AODV attains 60% of attack detection, SETORD attains 75% and Table 5 compares the suggested method for attack detection with the existing methods using percentages. In contrast to AODV's attack detection rate of 60%, SETORD's attack detection rate of 75%, and HEDAR's attack detection rate of 79%, the suggested method achieves 95% attack detection in black hole attacks, according to the graph.

Table 5. Comparison between the prior methods and proposed method for Attack detection (%).

| Methodologies / Attack | AODV (%) | SETORD (%) | HEDAR (%) | Proposed (%) |
|---|---|---|---|---|
| Black hole | 60 | 75 | 79 | 95 |
| Grey hole | 63 | 74 | 82 | 97 |
| Warm hole | 68 | 79 | 87 | 92 |
| Sybil | 65 | 69 | 84 | 94 |
| Altered/Replayed or Spoofed | 69 | 72 | 88 | 98 |
| Sinkhole | 64 | 74 | 78 | 96 |
| Hello flood | 66 | 76 | 85 | 94 |
| DOS attack-Apache2 | 62 | 71 | 89 | 98 |
| R2l | 65 | 69 | 83 | 93 |
| Probe | 69 | 73 | 89 | 97 |

In comparison to AODV, SETORD, and HEDAR, which each achieve 63, 74, and 82 percent of attack detection in Grey hole attacks, the suggested technique achieves 97 percent attack detection. In comparison to AODV's attack detection rate of 60%, SETORD's attack detection rate of 75%, and HEDAR's attack detection rate of 79%, the suggested approach detects attacks in black hole attacks at a rate of 95%. Comparing the suggested method to AODV, SETORD, and HEDAR, which each achieve 68, 79, and 87 percent attack detection in warm hole attacks, the proposed method achieves 92 percent attack detection. Additionally, the suggested method detects attacks at a rate of 94% compared to AODV's attack detection rate of 65%, SETORD's attack detection rate of 69%, and HEDAR's attack detection rate of 84% in the Sybil assault.

The suggested method achieves 98 percent attack detection in altered/replayed or spoof attacks compared to AODV's 69 percent attack detection, SETORD's 72 percent attack detection, and HEDAR's 88 percent attack detection. The suggested method also achieves 96 percent attack detection in comparison to AODV's attack detection rate of 64 percent, SETORD's attack detection rate of 74 percent, and HEDAR's attack detection rate of 78 percent in sinkhole attacks. The suggested technique thus achieves 94 percent attack detection in the Hello Flood Attack compared to AODV's 66 percent attack detection, SETORD's 76 percent assault detection, and HEDAR's 85 percent attack detection. As opposed to AODV's attack detection rate of 62%, SETORD's attack detection rate of 71%, and HEDAR's attack detection rate of 89%, the suggested technique detects attacks against Apache2 with 98 percent accuracy. The proposed technique thus achieves 93 percent attack detection in R21Attack compared to AODV's 65 percent attack detection, SETORD's 69 percent attack detection, and HEDAR's 83 percent attack detection. The suggested method thus achieves 97 percent attack detection in Probe Attack compared to AODV's attack detection rate of 69 percent, SETORD's attack detection rate of 73 percent, and HEDAR's attack detection rate of 89 percent. As a result, the suggested system will be more effective than both existing systems and those that do not use blockchain.

## 5. CONCLUSIONS

SRA with blockchain improves MANET security. Insufficient network integrity and dependence management cause the delay. SRA recommends using a Blockchain-Based Integrity System to improve security. Implement Promulgated Reliance Esteemed Quadruplets Condition in blockchain technology to boost system stability and scalability. Blockchain connects more networks to infect QoS caused by delay. The blockchain-based SAR will reduce latency. The findings show that the proposed framework improves several parameters, such as a high packet delivery ratio, a high throughput, a reduced number of dropped packets, and a minimised end-to-end delay, thereby improving the effectiveness of security in MANET through the use of blockchain technology and the subsequent transmission of data based on performance metrics. We will continue investigating blockchain and its application in mobile ad hoc networks in the upcoming work. We want to learn more about the potential security risk and assaults that blockchain in MANET thwarts. More research on hash consensus methods, malicious node detection, optimal routing patterns, and improvement between two nodes could be done to enhance this design. Additional studies in the fields of content creation, contribution, and transmission are also included.

### CONFLICTS OF INTEREST

The authors declare no conflict of interest.

### ACKNOWLEDGMENTS

I like to thank Dr. Raj Thaneeghavl. V., my primary supervisor, for guiding me through this study. In addition, also I would like to thank all authors for their contributions and the success of this manuscript and all editors and anonymous reviewers of this manuscript.

# AUTHORS

**Nitesh Ghodich*or*** is a Research Scholar at the Department of Computer Science & Engineering, SRK University, Bhopal. Master of Technology in Computer science & Engineering (2010). Bachelor of Engineering in Computer Technology (2001). after Engineering has been working as a Lecturer at Nagpur university. I have published more than 10 papers in Journal as an author and co-author with citation 28 and h-index 3.

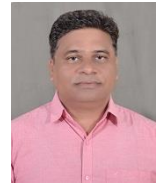

**Raj Thaneeghavl. V**, Professor at the Department of Computer Science, SRK University, Bhopal. He has completed Ph.D. in Computer Science in 2014, MCA in 2010, CCNA certified, Engineering degree earned in 2005. Research Activities in a different area and also research work published in more than 3 journals, more than 10 presentations.

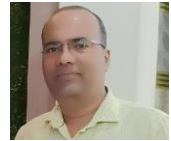

**Gautam M. Borkar** received his Bachelor's degree from the National Institute of Technology, jalandhar, Punjab, India and completed his Master's from the Sant Gadge Baba Amravati University, Amravati. Currently, he is working as an Assistant Professor in Rajiv Gandhi Institute of Technology, Mumbai and completed his PhD from the Sant Gadge Baba Amravati University, Amravati. His current research interests include network security, trust management and security in wireless sensor network.

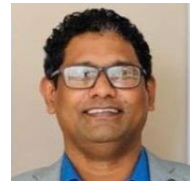

**Ankush Sawarkar** a Ph.D scholar and I am very close to finishing my Ph.D. thesis at Visvesvaraya National Institute of Technology (VNIT), Nagpur, in the Department of Computer Science and Engineering. My broad areas of interests include Bamboo, Artificial Intelligence, Machine Learning, Deep Learning, I did my Masters in Computer Network and Information Security from Shri Guru Gobind Singhji Institute of Engineering and Technology (SGGSIE&T) Nanded.

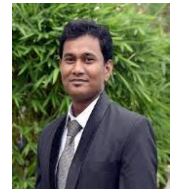